\documentclass[aps,prl, twocolumn,superscriptaddress,groupedaddress]{revtex4}
\usepackage{amssymb}
\usepackage{amsmath}
\usepackage{graphicx}
\usepackage{epstopdf}
\usepackage{bm}
\usepackage{gensymb}
\usepackage{soul}
\usepackage{sidecap}
\usepackage[normalem]{ulem}

\usepackage{ mathrsfs }
\usepackage {times}
\usepackage{float}
\usepackage[caption = false]{subfig}
\usepackage{subfig}
\usepackage{enumerate}
\usepackage{multirow}
\usepackage{tabularx}
\usepackage{array}
\usepackage{slantsc}
\usepackage{lmodern}
\newcommand\redout{\bgroup\markoverwith{\textcolor{red}{\rule[.5ex]{2pt}{0.4pt}}}\ULon}

\usepackage[colorlinks=true,citecolor=blue]{hyperref}
\hypersetup{colorlinks=true,citecolor=blue,linkcolor=red,urlcolor=blue}

\usepackage{hyperref,cleveref}
\graphicspath{{.}{./Figs/}}

\usepackage{color}
\usepackage[dvipsnames]{xcolor}
\usepackage{xcolor}

\begin{document}
\title{Driving enhanced quantum sensing in partially accessible many-body systems}


\author{Utkarsh Mishra}
\author{Abolfazl Bayat}
\email{abolfazl.bayat@uestc.edu.cn}
\affiliation{Institute of Fundamental and Frontier Sciences, University of Electronic Science and Technology of China, Chengdu 610054, China}

\begin{abstract}
The Ground-state criticality of many-body systems is a resource for quantum-enhanced sensing, namely the Heisenberg precision limit, provided that one has access to the whole system. We show that for partial accessibility, the sensing capabilities of a block of spins in the ground state reduces to the sub-Heisenberg limit. To compensate for this, we drive the hamiltonian periodically and use a local steady-state for quantum sensing. Remarkably, the steady-state sensing shows a significant enhancement in precision compared to the ground state and even achieves super-Heisenberg scaling for low frequencies. The origin of this precision enhancement is related to the closing of the Floquet quasienergy gap. It is in close correspondence with the vanishing of the energy gap at criticality for ground state sensing with global accessibility. The proposal is general to all the integrable models and can be implemented on existing quantum devices. 
\end{abstract}
\date{\today}
\maketitle

\emph{Introduction.--} The high sensitivity of quantum systems to  variations of their environment makes them 
superior sensors to their classical counterparts~\cite{Lewenstein_2016,Degen2017,RMP_2018,Nonclassical,Yip2019,Sensing_2019,
BioMed_2020,Mag_Spect_2019}. This is reflected in the Cram\'{e}r-Rao inequality, which determines the precision limit of estimating an unknown field $h$, quantified by the standard deviation $\delta h$, through $\delta h {\ge} 1/\sqrt{MF}$, where $M$ is the number of samples and $F$ is the Fisher information~\cite{caves_1994, Parish_2009}. While the classical Fisher information scales as $F_C{\sim} N$ (standard limit), with $N$ being the number of resources (e.g.,  number of particles) in the sensor, the quantum mechanics allows to go beyond this and achieve $F_Q{\sim} N^2$ (Heisenberg limit). Several quantum features are known to provide enhanced sensing precision: (i) entanglement in the special form of GHZ~\cite{Maccone_etal_2004,Nonclassical-1,
Maccone_etal_2006,GHZ,HL} or N00N~\cite{Coherent_state,N00N_state, N00N_state_1} states; (ii) wave function collapse resulted from sequential measurements separated by intervals of free evolution~\cite{sep_state_met,sep_state_met_1,sep_state_met_2,sep_state_met_3,
sep_state_met_4,sep_state_met_5,sep_state_met_6}; 
and (iii) quantum criticality in many-body systems~\cite{Zanardi2006,Zanardi2007,Zanardi_2010,
Paris1,SHIJIAN2010,QC_met,Kmolmer_2011}. Any of these approaches have their advantages and disadvantages. If a $d$-dimensional many-body system operates near its critical ground state, the quantum Fisher information (QFI) of the \emph{whole} system scales as $F_Q{\sim} N^{2/d\nu}$, where the $\nu$ characterizes the critical exponent for the divergence of the correlation length~\cite{Rams2018}.   
In the absence of global accessibility, one can only control a subsystem, which in general is a mixed state. A key question is: how does QFI scales with the subsystem size in a critical system? Besides, can the Heisenberg scaling be retrieved if the scaling becomes sub-Heisenberg, due to the mixedness of the subsystem?

Non-equilibrium dynamics of periodically driven many-body systems has been exploited for investigating the emergence of steady-state~\cite{Pd_steady_state1}, time-crystals~\cite{Floquet_crystal}, topological systems~\cite{top1,top2}, entanglement generation~\cite{ent1,ent22,ent33,ent4,ent5}, Floquet spectroscopy~\cite{spect1,spect2}, dynamically controlled quantum thermometry~\cite{VM_2019}, and dynamical phase transitions~\cite{FLoquet_dqpt,FLoquet_dqpt_1,FLoquet_dqpt_2}. The useful features of periodically driven many-body systems are: (i) any local subsystem reaches a steady-state; and (ii) the Floquet mechanism is applicable which simplifies the study of the dynamics. In non-integrable systems, a periodic field drives any small subsystem to a featureless infinite temperature thermal steady-state with no memory of the Hamiltonian parameters~\cite{Rigol_2019}. On the other hand, for integrable models, a non-trivial steady-state can be obtained that carries information about the Hamiltonian parameters~\cite{Heating,Pd_steady_state1,ent1,ent22,ent33,ent4,ent5,Brydges2019,UMAB}. An important, yet unexplored, open question is whether the local steady-states of periodically driven integrable systems can be used for enhancing the sensing precision in many-body sensors with partial accessibility.

In this paper, we address the above open problems by considering an $XY$ spin chain for detecting a transverse magnetic field. 
We first find that in the absence of global accessibility, the sensing precision, even at the critical point, diminishes to sub-Heisenberg scaling. Then, we show that by applying a proper periodic transverse field and exploiting the local steady-states we can even achieve super-Heisenberg sensitivity. Remarkably, this enhanced sensing is not limited to the critical points of the system and exists for all the points across the phase diagram with a vanishing Floquet quasienergy gap.  The protocol can be realized in existing quantum devices using simple measurements.


\emph{Model.--} We consider quantum $XY$ spin chain for measuring an unknown static transverse magnetic field $h_0$. To manipulate the system for the desired accuracy we apply a periodic transverse field, $h(t)$, 
to the system. Therefore, the total Hamiltonian can be written as
\vspace{-0.2cm }
\begin{eqnarray}
    H (t)&=& -\frac{J}{2}\sum_{i=1}^{N}  \Big[\Big(\frac{1+\gamma}{2}\Big)\sigma^{x}_{i}\sigma^{x}_{i+1}+\Big(\frac{1-\gamma}{2}\Big)\sigma^{y}_{i}\sigma^{y}_{i+1}\Big]\nonumber\\
    &-& \frac{(h_0+h(t))}{2} \sum_{i}\sigma^{z}_{i},
\label{eq:model}
\end{eqnarray}
where, $\sigma^{\Theta}$($\Theta = x,y,z$) are the Pauli matrices, $J$ (which is set to be 1 throughout the paper) is the exchange coupling,   $-1{\leq} \gamma {\leq} 1$ is the anisotropic parameter, and the periodic-boundary conditions, i.e., $\sigma^{\alpha}_{N+1} {\equiv} \sigma^{\alpha}_{1}$, is imposed. At time $t{=}0$, a periodic field of the form of $h(t){=}h_1\sin(\omega t)$ is applied to the system, where $\omega=2\pi/\tau$ with $\tau$ being the time-period. The Hamiltonian $H(0)$ 
shows quantum criticality at $h_0{=}h_c$ such that $h_c{/}J{=}1$ for all values of $\gamma$~\cite{Sachdev2017}. We consider that system is initially prepared in the ground state of $H(0)$. However, as discussed in the Supplementary Materials (SM) \cite{SM}, the proposed mechanism is general and works for other initial states. By switching the probe field $h(t)$ the initial state starts to evolve. The exact solution for the evolved is provided in the SM.

\emph{Sensing with global accessibility.--} If one has access to the whole system, namely $|\Psi_{0}(t)\rangle$, then the QFI is given by $F_Q(t){=}4\chi_F(t)$, where $\chi_F(t){=}\langle \partial_{h_0}\Psi(t)|\partial_{h_0}\Psi(t)\rangle{-}|\langle \Psi(t)|\partial_{h_0}\Psi(t)\rangle|^2$.
Especially for $H(0)$ the global QFI has been extensively studied and it was shown that at the ground state criticality it scales as $F_Q^{gs}{=}F_Q(0){\sim} N^2$~\cite{Rams2018,Zanardi2006,Zanardi2007,Zanardi_2010,
Paris1,SHIJIAN2010,QC_met,scaling_fd,scaling_fd_1,scaling_fd_2,
scaling_fd_3,scaling_fd_4,scaling_fd_5,scaling_fd_6} while away from the criticality it scales as $F_Q^{gs}{=}F_Q(0){\sim} N$.  We show this in the SM by simulating the QFI of the global system. In the rest of the letter, we focus on partial accessibility.

\emph{Sensing with partial accessibility.--} In the absence of global accessibility, one has to rely on accessing a local block of size $L$ with $L\ll N$. The partially accessible state of the system is described by the reduced density matrix obtained by tracing out all particles out of the block $L$, namely 
$\rho_L(t){=}\mbox{tr}_{N-L}\Big(|\Psi_{0}(t)\rangle \langle \Psi_{0}(t)|\Big).$ The QFI of the state is given by~\cite{Parish_2009}
\begin{equation}
F_{Q}=\sum_{r,s=1}^{2^L}\frac{2 {\Re}(\langle \lambda_{r}|\partial_{h_0}\rho_{L}|\lambda_{s}\rangle \langle \lambda_{s}|\partial_{h_0}\rho_{L}|\lambda_{r}\rangle)}{\lambda_{r}+\lambda_s},
\label{eq:QFI} 
\end{equation}
where, $\rho_L = \sum_{r=1}^{2^L}\lambda_{r}|\lambda_{r}\rangle \langle \lambda_{r}|$ with $\lambda_r$ and $|\lambda_{r}\rangle$ being the eigenvalues and eigenvectors of $\rho_L$, respectively. ${\Re}[\cdot]$ denotes the real parts of the quantity inside the  parenthesis and the sum  excludes terms for which $\lambda_r+\lambda_s=0$.  Note that the QFI is independent of the choice of the measurement operators and, in general,  depends on the unknown parameter $h_0$. Calculation of the QFI for the state $\rho_L$ is given in the SM

%

\emph{Steady-state of a block.--} After a long-time $t$, the reduced density matrix $\rho_{L}(t)$ equilibriates to a steady-state. Our goal in this paper is to measure the  QFI for such a steady-state. By using $H(t+\tau){=}H(t)$ and Floquet formalism, one can  obtain the time-evolved state after $n{-}$cycles from an initial state $|\Psi_0\rangle$ as $|\Psi(n\tau)\rangle{=}\sum_{i}e^{-{\it i}\mu_{i}n\tau}|\mu_{i}\rangle\langle\mu_{i}|\Psi_0\rangle$. Here $\{\mu_{i},|\mu_{i}\rangle\}$ are the eigenvalues (Floquet quasienergies) and eigenvectors of the one-period Floquet operator $U(\tau){=}{\cal T}e^{-{\it i}\int_{0}^{\tau}H(t)dt}$, with ${\cal T}$ being time-order operator. The expectation value of a local operator, ${\cal O}$, in the time-evolved state then can be expressed as $\langle {\cal O}\rangle{=}\sum_{\ell}\langle\mu_{\ell}|{\cal O}|\mu_{\ell}\rangle|\langle\mu_{\ell}|\Psi_{0}\rangle|^2+\sum_{\ell \neq j}\langle\mu_{\ell}|{\cal O}|\mu_{j}\rangle\langle\mu_{\ell}|\Psi_{0}\rangle\langle\Psi_{0}|\mu_{j}\rangle e^{-2{\it i}(\mu_{\ell}-\mu_{j})n\tau}$. 
The first and second terms describe the diagonal contribution and the fluctuation around the diagonal term, respectively. The second term vanishes for a long time (Riemann-Lebesgue lemma). Using the above formalism, we calculate the expectation value of the fermionic correlation functions in the limit $t\to \infty$. (see SM for obtaining the local steady-state of the model).
These correlation functions give the steady-state QFI, namely $F^{ss}_{Q}{=}\lim\limits_{t\rightarrow \infty} F_Q(t)$ for the state $\rho_L$. 
\begin{figure}
\includegraphics[width=0.45\textwidth]{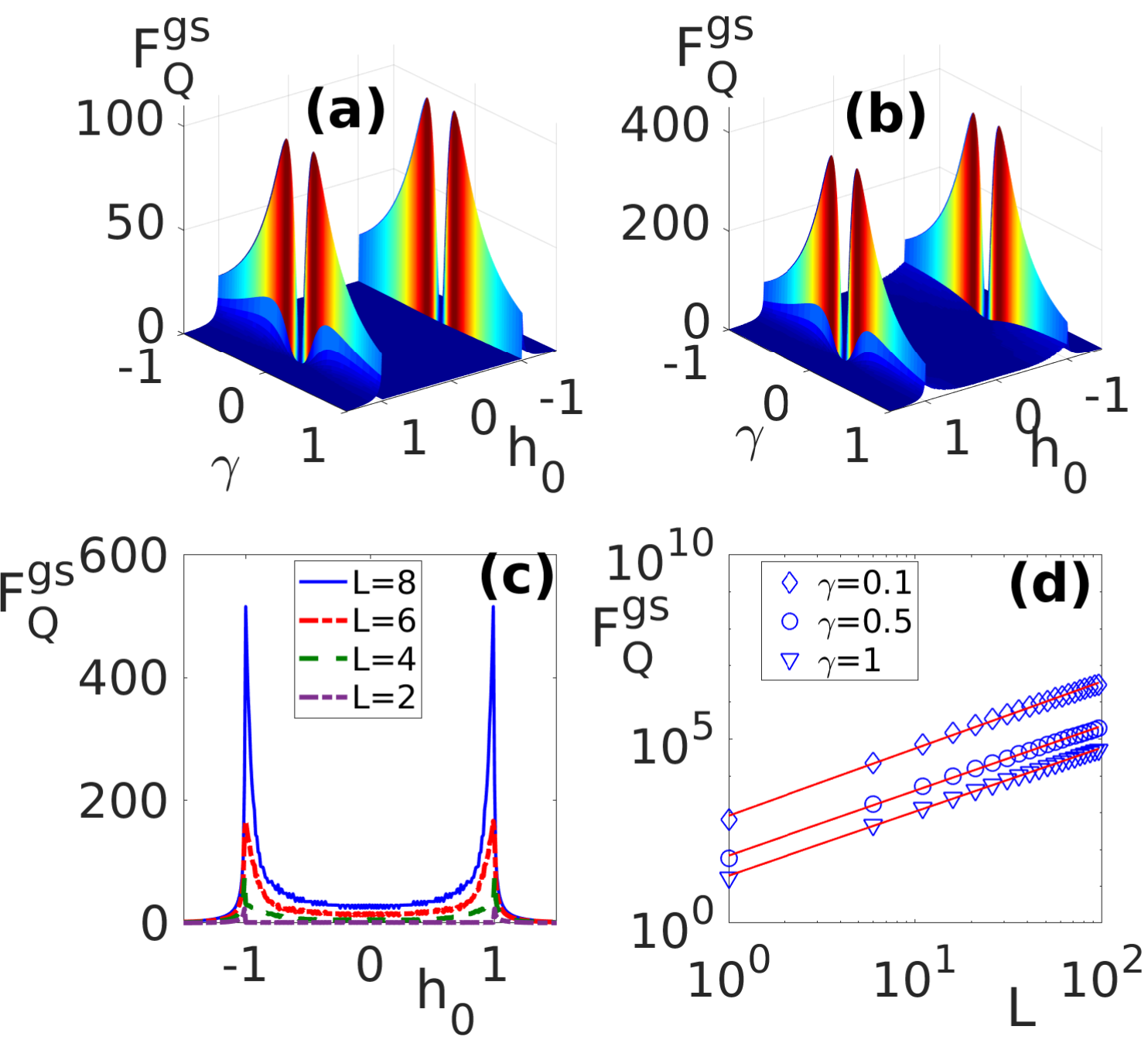}
\caption{ The QFI of the ground state, $F^{gs}_{Q}$, as a function of $\gamma$ and $h_0$ for (a) $L{=}2$ and (b) $L{=}4$. (c) The  $F^{gs}_Q$ as a function of $h_0$ when $\gamma{=}1$ for various choices of $L$. (d) The log-log scaling of $F^{gs}_Q$ as a function of $L$ for different $\gamma$'s.   The fitting is shown by regular line whereas the markers represent the original data. In all panels $N{=}6000$.}
\label{fig:fig1}
\end{figure}

\emph{Ground state sensing.--} 
In the absence of global accessibility, one has to rely on the sensing capability of $\rho_L$, which in general is a mixed state. This mixedness can diminish the sensing capability. To quantify this, we consider the ground state of $H(0)$ for $N{=}6000$ and plot the QFI, namely $F^{gs}_Q$, for $L{=}2$ and $L{=}4$ in Figs.~\ref{fig:fig1}(a) and (b), respectively. It can be seen from the plots that $F^{gs}_Q$ shows peaks at points $h_0/J{=}\pm 1$ that marks the quantum criticality of the system. It is an interesting observation that not only the QFI of a full chain but also that of the reduced state distinguishes the criticality~\cite{Zanardi2006,SHIJIAN2010,Sacramento_2011,Park_2016,Yu2016}. In Figs.~\ref{fig:fig1}(a) and (b), the $F^{gs}_Q$ becomes vanishingly small at $\gamma{=}0$.  
Since for  $\gamma{=}0$, the field part of $H(0)$ commutes with the interaction part, the variation of the field $h_0$ does not induce any change in the ground state of $H(0)$ which reflects itself in $F^{gs}_Q{=}0$. To have a better understanding of the role of  $L$, we plot $F^{gs}_Q$ 
versus $h_0$  at $\gamma{=}1$ in Fig.~\ref{fig:fig1}(c)
for various $L$'s. The QFI increases with $L$ and this effect becomes even more pronounced  at the critical point $h_0/J=\pm 1$. To have a quantitative analysis for the scaling of the QFI at the critical point, in Fig.~\ref{fig:fig1}(d) we plot $F^{gs}_Q$ as a function of $L$ for $\gamma{=}0.1,0.5,1$ by fixing $h_0/J{=}1$. The scaling follows a power-law form, i.e., $F_Q^{gs}(h{=}h_c) \sim a L^{\eta}$. Numerical fitting results in $(a,\eta){=}(6.718,1.8)$ for $\gamma{=}0.1$, $(a,\eta){=}(4.2235,1.76)$ for $\gamma{=}0.5$, and $(a,\eta){=}(2.967,1.74)$ for $\gamma{=}1$, respectively. Thus, for the critical ground state and with partial accessibility, the QFI scales weaker than the Heisenberg bound (i.e., $\eta{=}2$), although it still outperforms the standard limit (i.e., $\eta{=}1$) showing quantum-enhanced sensing. Is it possible to improve this and retrieve Heisenberg scaling? 

\begin{figure}
\includegraphics[width=0.5\textwidth]{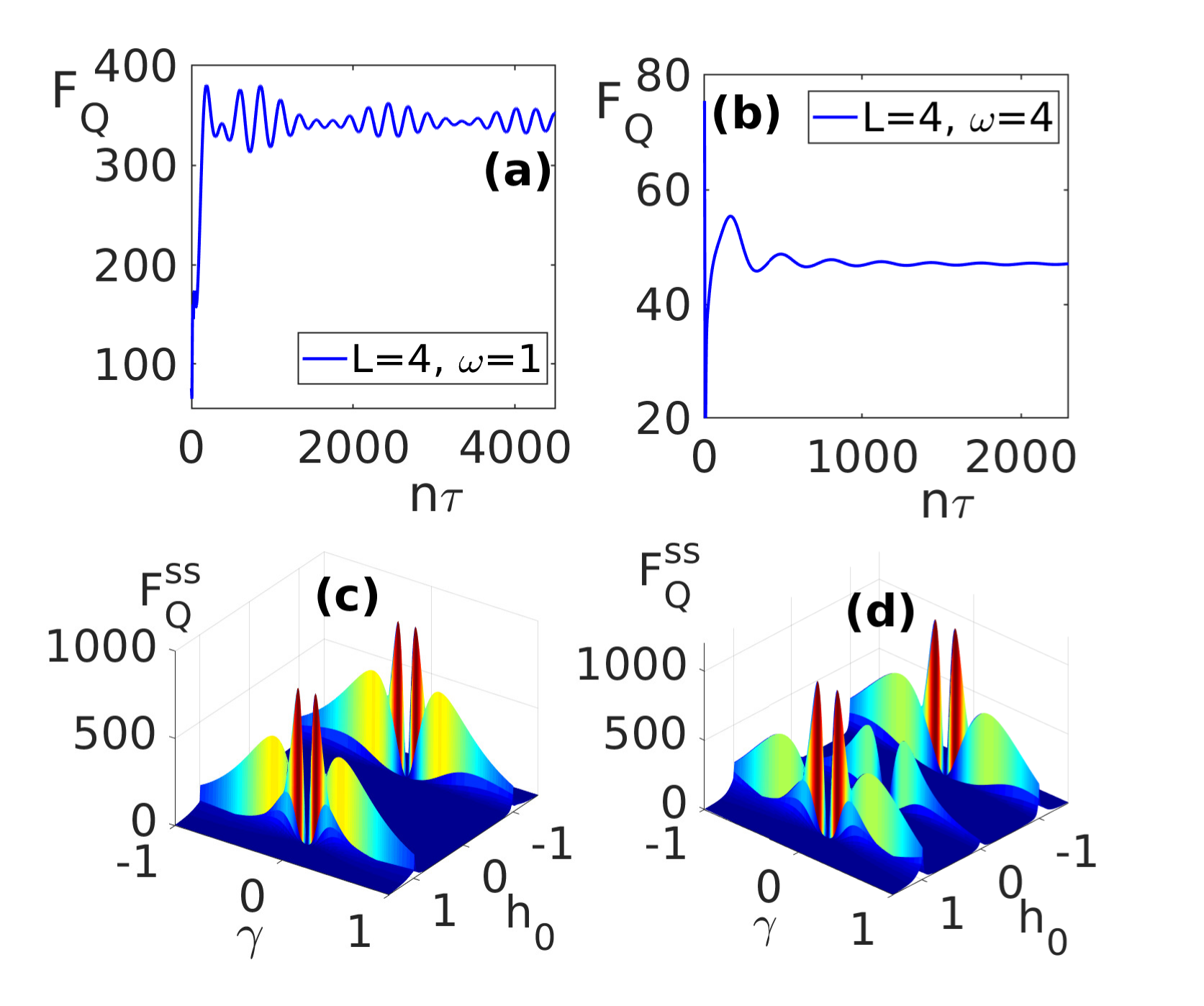}
\caption{$F_Q$ as a function of time $t{=}n\tau$ for (a) $\omega{=}1$ (b) $\omega{=}4$, for $L{=}4$, $h_0/J{=}1$, and $\gamma=1$. Steady-state  $F^{ss}_{Q}$ as a function of $h_0$ and $\gamma$ for (c) $\omega{=}4$ and (d) $\omega{=}2$.  In all panels, the system size is $N{=}6000$ and $h_1=1.5$. }
\label{fig:fig2}
\end{figure} 

\emph{Steady-state sensing.--} To enhance the sensing capability with $\rho_L(t)$, we propose to apply a periodic drive as given in Eq.~(\ref{eq:model}). The resulting dynamics tend to thermalize the quantum state of the block.   
 In non-integrable systems, while the global quantum state can still be used for quantum sensing~\cite{ref1,ref2}, the subsystems equilibrate to an infinite temperature state, and carry no information about the Hamiltonian~\cite{Rigol_2019}. In integrable models, as in Eq.~(\ref{eq:model}), the steady-state does \emph{not} thermalize to the infinite temperature due to local conserved quantities, and thus, carries a wealth of information about the parameters of the system~\cite{Pd_steady_state1}. To find the sensing capability of the steady-state of a block of $L=4$, in Figs.~\ref{fig:fig2}(a)-(b) we plot  $F_Q(n\tau)$ as a function of time $t{=}n\tau$ for $\omega{=}1,4$, respectively. The QFI reaches an equilibrium after a short transition time.
Equilibration of the probe state is of multifold importance: (i) the imprinted information of $h_0$ in the density matrix may enhance the sensitivity and (ii) the emergent steady-state remains almost fixed in time which simplifies the measurement.

To see the sensing capability of the steady-state for a choice of $h_1{=}1.5$ and $L=4$, we compute the steady-state QFI, denoted as $F^{ss}_{Q}$. In Fig.~\ref{fig:fig2}(c), we plot
$F_Q^{ss}$ as a function of $\gamma$ and $h_0$ for $\omega{=}4$.
The $F^{ss}_{Q}$ shows similar behavior as $F^{gs}_{Q}$ in Figs.~\ref{fig:fig1}(a)-(b), except around $\gamma{=}0$ (the behavior  of $F^{ss}_{Q}$ as a function of $\gamma$ is discussed in the SM). 
In Fig.~\ref{fig:fig2}(d), we plot the $F^{ss}_{Q}$ for a lower frequency ($\omega{=}2$). Interestingly the $F^{ss}_{Q}$ becomes non-zero along the line $h_0{=}0$, whereas it is zero for $\omega{=}4$.  Thus, by properly driving the system, extra peaks appear in the QFI even away from the ground state criticality and thus achieve quantum-enhanced sensing over a wider range.

\begin{figure}
\includegraphics[width=0.45\textwidth]{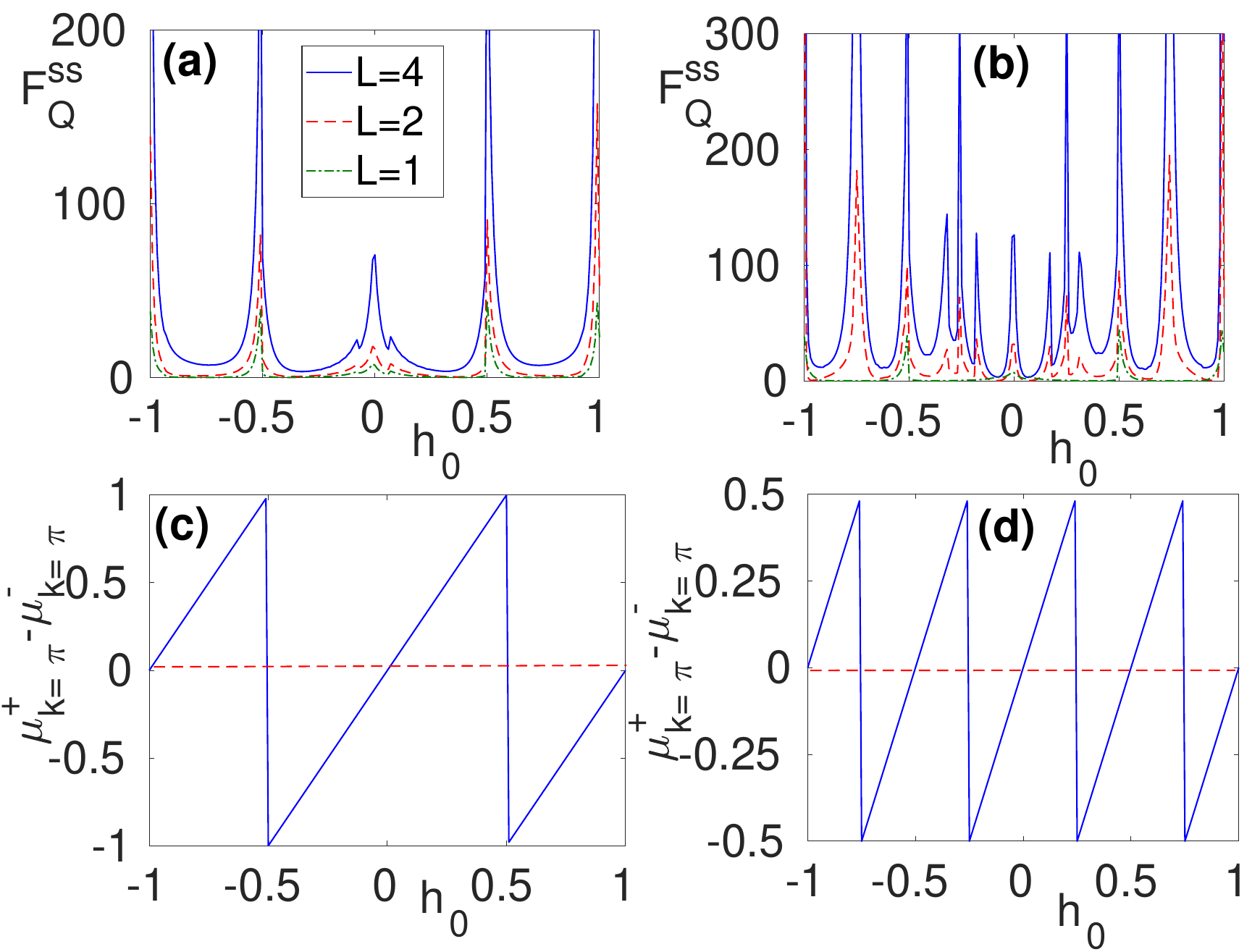}
\caption{The QFI in the steady-state with respect to $h_0$ for different frequency: (a) $\omega{=}1$; and (b) $\omega{=}0.5$. The difference of Floquet quasienergies $\mu^{\pm}_{k=\pi}$ for frequencies: (c) $\omega{=}1$, and (d) $\omega{=}0.5$.
In all panels, $N{=}6000$, $h_1{=}1.5$ and $\gamma{=}1$. }
\label{fig:fig3}
\end{figure}

\begin{figure}
\includegraphics[width=0.45\textwidth]{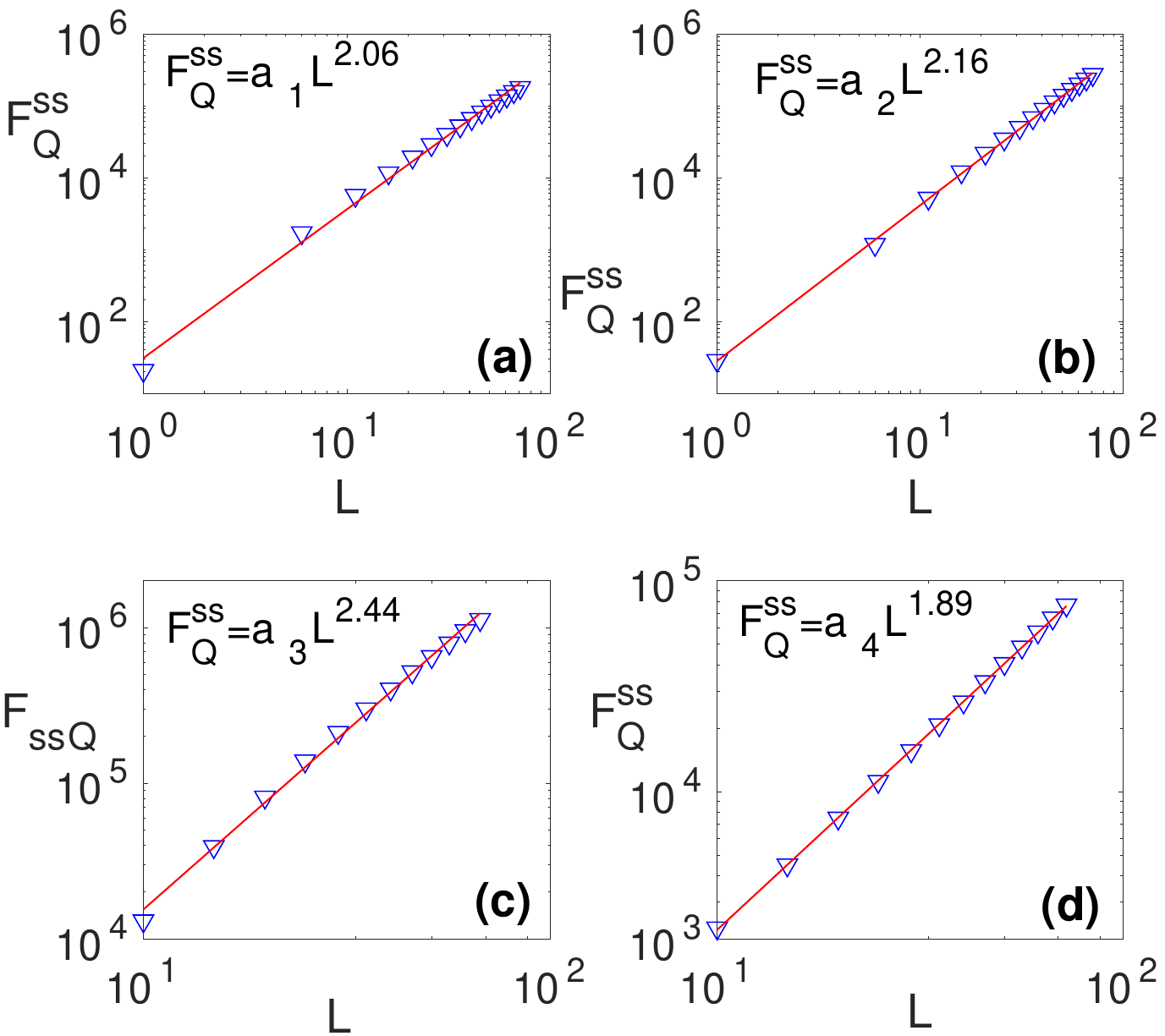}
\caption{Scaling of the $F^{ss}_{Q}$ vs $L$ along the vanishing Floquet quasienergy gap line for: (a) $(\omega,h_0){=}(2,1)$; (b) $(\omega,h_0){=}(1,1)$; (c) $(\omega,h_0){=}(0.5,1)$; and (d) $(\omega,h_0){=}(1,0.5)$. The coefficients are: $a_1{=}3.43$, $a_2{=}3.33$, $a_3=3.78$, and $a_4=3.15$.  In all panels, $\gamma{=}1$, $h_1{=}1.5$.}
\label{fig:fig4}
\end{figure}

\emph{Floquet resonance.--} To investigate the emergence of extra peaks, we fix the parameters $\gamma{=}1$,  $h_1{=}1$, and plot the $F^{ss}_{Q}$ as a function of $h_0$  in Figs.~\ref{fig:fig3}(a)-(b) for frequencies $\omega{=}1$ and  $\omega{=}0.5$, respectively. In each panel, the different curves are for different block size $L$. It can be seen clearly from the plots that the number of peaks increases as the frequency gets smaller. The peaks are related to the eigenvalues of the one-period Floquet operator $U_k(\tau)$, where $U_k(\tau)$ is the Floquet operator for each quasimomentum mode $k\in[0,\pi]$, as discussed in the SM. 
The eigenvalues of $U_{k}(\tau)$ can be written as $e^{i\tau \mu_{k}^{\pm}}$ where $\mu^{\pm}_{k} = \pm\frac{\omega}{\pi}\tan^{-1}\sqrt{\frac{1-\Re(u_{k}(\tau))}{1+\Re(u_{k}(\tau))}}$,  are the Floquet quasienergies~\cite{Pd_steady_state1}. Interestingly, the peaks occur at the position of Floquet resonances, i.e., when $\mu_{k}^{+}{=}\mu_{k}^{-}$. 
For $h_1 \neq 0$,  the quasienergy spectrum shows avoided crossing except at $k=0$ and $k=\pi$. Thus, the Floquet resonance condition will only be satisfied by modes at $k=0$ and $k=\pi$.  
Therefore, the Floquet resonance condition for the energy eigenvalues becomes $2E_{k=0,\pi}(t=0)= q \omega$ for some integer $q$, where $E_{k}(t=0)=\pm \sqrt{(h_0-J\cos(k))^2+J^2\gamma^2\sin(k)^2}$ (see the SM for definition of $E_k$).
We depict the behavior of Floquet quasienergy gap, i.e., $\mu^{+}_{k=\pi}-\mu^{-}_{k=\pi}$ as a function of $h_0$ for $\omega=1$ and $\omega=0.5$ in Figs.~\ref{fig:fig3}(c)-(d), respectively. It can be seen that for each peak in Figs.~\ref{fig:fig3}(a)-(b), the quasienergy gap vanishes at those $h_0$. Thus, the vanishing of the quasienergy gap is responsible for the peaks in the $F^{ss}_{Q}$ observed in Figs.~\ref{fig:fig3}(a)-(b). The detailed calculation of Floquet formalism and quasienergy gap is provided in the SM.

\begin{figure}
\includegraphics[width=0.45\textwidth]{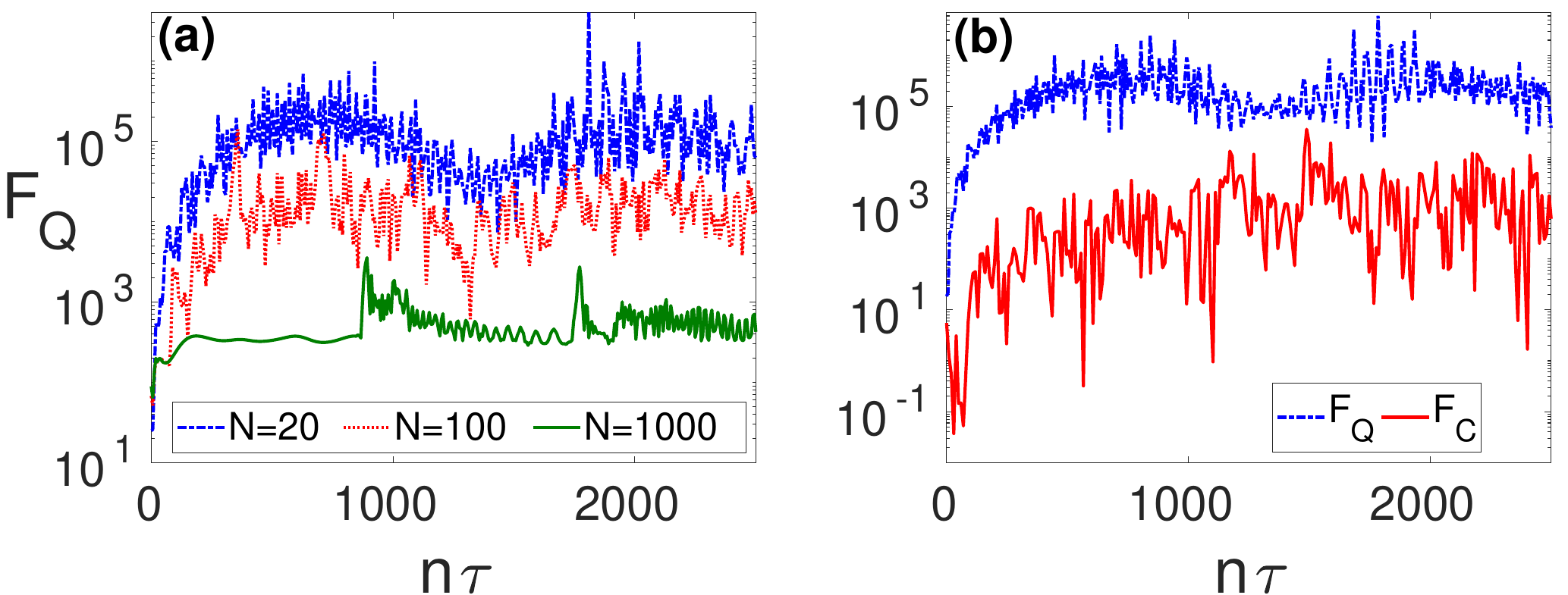}
\caption{(a) Dynamics of QFI for different system size $N$ and fixed block-size $L{=}4$. (b) Dynamics of both quantum and classical Fisher information of a block of $L{=}4$ for a system of size $N{=}14$.  In all panels,  $h_0{=}1, \gamma{=}1$, $h_1{=}1.5$ and $\omega{=}1$. }
\label{fig:fig5}
\end{figure}

\emph{Driving enhanced sensing.--} As seen above, driving the system can enhance the steady-state QFI. It is of utmost interest to see whether this can improve the scaling of the QFI as a function of $L$. We first focus on the critical point, i.e. $h_0/J{=}1$, and without loss of generality fix the parameters $\gamma{=}1$ and $h_1{=}1.5$. In Figs.~\ref{fig:fig4}(a)-(c) we plot $F^{ss}_Q$ versus $L$ together with a power-law fitting function $\widetilde{F}^{ss}_Q \sim aL^{\eta}$ at $h_0/J=1$ for different frequencies such that $F^{ss}_{Q}\approx \widetilde{F}^{ss}_{Q}$. 
The coefficient $\eta$ shows that in the range $\omega \leq 2$ the scaling of the steady-state surpasses the scaling of the ground state. Remarkably, by tuning the driving frequency to $\omega{=}2$, see Fig.~\ref{fig:fig4}(a), one can indeed retrieve the Heisenberg scaling. Further decreasing the frequency can lead to the remarkable super-Heisenberg scaling of $\eta>2$, showing in Figs.~\ref{fig:fig4}(b) and (c).
This driving enhanced sensitivity is not limited to the critical point. In Fig.~\ref{fig:fig4}(d) we depict the scaling of the QFI versus the block size $L$ for $\omega{=}1$ at $h_0/J{=}0.5$, where the $F_Q^{ss}$ peaks  due to Floquet resonance, see Fig.~\ref{fig:fig3}(a). Interestingly, the scaling ($\eta{=}1.8$) exceeds the standard quantum limit showing that quantum-enhanced sensing can be achieved at all Floquet resonances. 


\emph{Role of the frequency.--} As discussed earlier, the enhanced precision is directly related to the vanishing quasienergy gap, which is a function of the frequency of the driving field. In fact, the frequency $\omega$ has two roles. First, for $\omega{<}2$, the QFI shows extra peaks which are absent in the phase diagram of the ground state, e.g. see Fig.~\ref{fig:fig2}(d). Second, at the vanishing Floquet quasienergy gap points, lowering the $\omega$ results in better scaling. For instance, as shown in Figs.~\ref{fig:fig4}(a)-(c), for the critical field $h_0{=}1$ one can achieve super-Heisenberg scaling once $\omega{<}1$.

\emph{Realization in near-term quantum simulators.--} Among the emerging quantum simulators ion-traps~\cite{CMonroe_ion_trap, P_Zoller, R_Blatt} and superconducting devices~\cite{Roushan_MBL,Guo_MBL,Ming_MBL} are the best candidates for the realization of our protocol as their interaction can be described by the Hamiltonian in Eq.~(\ref{eq:model}). Near-term quantum devices and simulators are limited in size~\cite{John_Preskill}. To investigate the performance of our protocol on small systems, in  Fig.~\ref{fig:fig5}(a) we plot $F_Q$ for a block of size $L{=}4$ as a function of time for various total system sizes. Interestingly, small systems provide high quantum Fisher information indicating more potential for sensing. It is because the larger the system, the more degrees of freedom for the dispersion of information.  

It is worth emphasizing that $F_Q$ provides an ultimate bound for sensing precision attained only if the measurement basis is optimal. However, the optimal measurement basis might be complicated and depends on the unknown parameter that makes the saturation of the Cram\'{e}r-Rao bound very challenging. Here, we consider a simple (but non-optimal) block magnetization measurement along the $x$ direction and compute its corresponding classical Fisher information $F_C$ (the definition of $F_C$ is in the SM). In  Fig.~\ref{fig:fig5}(b), we plot both  $F_C$ and $F_Q$ for a block of size $L{=}4$  as a function of time in a system of length $N{=}14$. Interestingly, the $F_C$ not only follows the behavior of $F_Q$ but also takes high values. It shows that a simple non-optimal measurement can serve for sensing. 

\emph{Conclusion.--} In this letter, we have shown that in the absence of global accessibility of the whole state, the Heisenberg scaling of the QFI for the critical many-body ground states of integrable systems reduces to sub-Heisenberg. To retrieve the Heisenberg scaling, we proposed to drive the system using a periodic field and use the steady-state of a block for sensing. Our results show that by tuning the frequency of the periodic field, one can generate multiple peaks across the phase diagram, improves the sensing over a larger interval. The scaling at all these peaks exceeds the standard limit precision and shows significant enhancement compared to the ground state.  Remarkably, at lower frequencies, one can even achieve super-Heisenberg scaling for the QFI. This steady-state quantum-enhanced sensitivity can be explained by the closing of the Floquet quasienergy gap. The protocol is general to all integrable models and best suited for ion traps and superconducting devices in which even a simple non-optimal measurement, such as block magnetization, can be used for achieving high precision.  

\emph{Acknowledgment.--}
AB thanks the National Key R\&D Program of China (Grant No.2018YFA0306703) and National Science Foundation of China (Grants No.12050410253 and No.92065115) for their support. UM acknowledges funding from the Chinese Postdoctoral Science Fund 2018M643437.


\begin{thebibliography}{10}
\bibitem{quasi} We obtained the quasienergies by diagonalizing $A = -1(\mathbb{I}-U(\tau))(\mathbb{I}+U(\tau))^{-1}$. The quasienergies are given by $\mu_{\alpha} =( \omega_0/\pi)\tan^{-1}(a_{\alpha})$, where $A = \sum_{\alpha}a_{\alpha}|a_{\alpha}\rangle\langle a_{\alpha}|$.
\end{thebibliography}

\clearpage
\onecolumngrid
\begin{center}
{\large \bf Supplementary Material for  \protect \\ 
``Driving enhanced quantum sensing in partially accessible many-body systems'' }\\
\vspace*{0.3cm}
Utkarsh Mishra$^{1}$ and Abolfazl Bayat$^{1}$ \\
$^{1}${\small \em Institute of Fundamental and Frontier Sciences, University of Electronic Science and Technology of China, Chengdu 610051, China} \\
\end{center}

\setcounter{equation}{0}
\setcounter{figure}{0}
\setcounter{table}{0}
\setcounter{page}{1}
\makeatletter
\renewcommand{\theequation}{S\arabic{equation}}
\renewcommand{\thefigure}{S\arabic{figure}}
\renewcommand{\bibnumfmt}[1]{[S#1]}
\renewcommand{\citenumfont}[1]{S#1}

\section{A. Diagonalization of the Hamiltonian}
The Hamiltonian in Eq.~(1) of the main text can be diagonalized even in the presence of time-varying field. We first decompose Pauli spin operators in terms of raising and lowering operators $\sigma^{\pm}$ as $\sigma^{x} =\sigma^{+}+ \sigma^{-}$ and $\sigma^{y} = -i(\sigma^{+}- \sigma^{-})$. We also obtain a transformation for $\sigma^{z}$ as $\sigma^{z} = 2\sigma^{+}\sigma^{-}-\mathbb{I}$. In terms of these operators, the Hamiltonian in Eq.~(1) can be written as 
\begin{eqnarray}
H(t) &{=}& -\frac{J}{2}\sum_{i=1}^{N}\Big(\gamma(\sigma_{i}^{+}\sigma_{i+1}^{+}+\sigma_{i+1}^{-}\sigma_{i}^{-})+(\sigma_{i}^{+}\sigma_{i+1}^{-}+\sigma_{i+1}^{-}\sigma_{i}^{+})\Big)\nonumber\\
&-&\sum_{i}\frac{(h_0+h(t))}{2}(2\sigma^{+}\sigma^{-}-\mathbb{I}).
\end{eqnarray}
The raising and lowering operators satisfy the anti-commutation relation, $\{\sigma_{i}^{-},\sigma_{j}^{+}\}=\delta_{i,j}$, with $(\sigma^{+}_{i})^2=0=(\sigma^{-}_{i})^2$. Thus, the operator $\sigma^{\pm}$ partly resemble the Fermi operators. Moreover, they also satisfy $[\sigma_{i}^{+},\sigma_{j}^{-}]=[\sigma_{i}^{+},\sigma_{j}^{+}]=[\sigma_{i}^{-},\sigma_{j}^{-}]=0$ for $i\neq j$. It can be noticed that the problem of diagonalizing $H$ remains intact as principle axis transformation of $\sigma^{\pm}$ does not lead to a proper set of operators which satisfy both commutation and anti-commutation relation. It is discovered~\cite{Lieb1961,Pfeuty1970} that a new set of Fermi operators can be defined in terms of which the Hamiltonian  
transforms into a simple form. This transformation, known as the Jordan-Wigner transformation, is defined as $\sigma^{-}_i = \exp[-i \sum_{\ell=1}^{i-1}\sigma^{+}_{\ell}\sigma^{-}_{\ell}]c_{i}$ and $\sigma^{+}_i = c^{\dagger}_{i}\exp[-i \sum_{\ell=1}^{i-1}\sigma^{+}_{\ell}\sigma^{-}_{\ell}]$. The $c^{\dagger}_{i}(c_{i})$ operators are Fermi operators $\{c_{i},c^{\dagger}_{j}\}=\delta_{ij}$, $\{c^{\dagger}_{i},c^{\dagger}_{j}\}=0=\{c_{i},c_{j}\}$. In terms of the new variables $c^{\dagger}_{i}(c_{i})$, the Hamiltonian becomes
\begin{eqnarray}
H(t) &=& -\frac{J}{2}\sum_{i=1}^{N}\Big(\gamma(c_{i}^{\dagger}c^{\dagger}_{i+1}+c_{i+1}c_{i})+(c_{i}^{\dagger}c_{i+1}+c^{\dagger}_{i+1}c_{i})\Big)\nonumber\\
&-&\sum_{i}\frac{(h_0+h(t))}{2}(2c^{\dagger}_{i}c_{i}-\mathbb{I}).
\end{eqnarray}
For the above Hamiltonian, one can define a parity operator ${\cal P}=\sum_{i}c^{\dagger}_{i}c_{i}$. The Hamiltonian commutes with ${\cal P}$ i.e., $[{\cal P},H]=0$. Thus, the Hamiltonian can be divided into any one of the parity sectors. We consider even system size and the positive parity sector. 
Now, it is customary to define Fourier transformation of the operators $c^{\dagger}_{i}(c_{i})$ as $c_{\ell} = \frac{1}{N}\sum_{k}e^{-ik\ell}d_{k}$ and similarly for $c^{\dagger}_{\ell}$, where $k=\frac{\pi}{N},\frac{3\pi}{N},\dots, \frac{(N-1)\pi}{N}$.  The new Hamiltonian can be written as 
\begin{eqnarray}
H(t) &=& -\frac{J}{2}\sum_{k>0}\Big(2\gamma\sin(k)(d_{k}^{\dagger}d^{\dagger}_{-k}+d_{-k}d_{k})\nonumber\\
&+&2 \cos(k)(d_{k}^{\dagger}d_{k}+d^{\dagger}_{-k}d_{-k})\Big)\nonumber\\
&-&\sum_{k}\frac{(h_0+h(t))}{2}(2d^{\dagger}_{k}d_{k}-\mathbb{I}).
\end{eqnarray}
By combining the terms and dropping the constant part from the Hamiltonian, we have 
\begin{eqnarray}
H(t) &=& -J\sum_{k>0}\gamma\sin(k)(d_{k}^{\dagger}d^{\dagger}_{-k}+d_{-k}d_{k})\nonumber\\
&-&\sum_{k>0}\Big(h_0+h(t)-J\cos(k)\Big)(d^{\dagger}_{k}d_{k}+d^{\dagger}_{-k}d_{-k}).
\label{eq:Hkt}
\end{eqnarray}
Thus, the full Hamiltonian is expressed as sum of Hamiltonian  $H_{k}(t)$ for each mode. By rotating $d_k$ and $d^{\dagger}_{k}$ as $d_k = \cos(\theta_k)B_k-\sin(\theta_k)B_{-k}$ with $\{B_k,B^{\dagger}_{k'}\}=\delta_{kk'}$, $\{B_{k},B_{k'}\}=0$, the Hamiltonian can be diagonalized at each instant $t$ as
\begin{eqnarray}
H (t)= \sum_{k}E_{k}(t)(2B^{\dagger}_{k}B_{k}-\mathbb{I}),
\end{eqnarray}
where, $E_{k}(t){=}\pm \sqrt{(h_0+h(t)-J\cos(k))^2+J^2\gamma^2(\sin(k)^2}$, is the instantaneous energy and $\theta_k = \tan^{-1}\frac{J\gamma \sin(k)}{h_0+h(t)-J\cos(k)}$. The ground state spectrum is given by $E_{k}(t=0)$. 

We consider that system is initially prepared in the ground state of $H(0)$ which is expressed as~\cite{Sachdev2017}
\begin{eqnarray}
|\Psi_{0}\rangle =\bigotimes_{k}\Big(v_k(0)|0\rangle + u_k(0)d^{\dagger}_{k}d^{\dagger}_{-k}|0\rangle\Big),
\label{eq:gs}
\end{eqnarray}
where, $u_k(0) {=} \sin(\theta_k/2)$ and $v_{k}(0){=}\cos(\theta_k/2)$ with $\theta_k{=}\tan^{-1}\frac{\gamma\sin(k)}{h_0-\cos(k)}$. 
Due to the presence of the probe field $h(t)$ the initial state evolves  to  $|\Psi(t)\rangle{=}U(t)|\Psi_{0}\rangle {=}\bigotimes_{k}\Big(v_k(t)|0\rangle + u_k(t)d^{\dagger}_{k}d^{\dagger}_{-k}|0\rangle\Big)$.
The propagator $U(t)$ is give by 
$U(t){=} {\cal T}e^{-i\int_{0}^{t}H(s)ds}$,
where ${\cal T}$ is the time-ordered product. Since all the $H_k$'s are commuting, we have  $U(t)=\bigotimes_k U_{k}(t)$,  where $U_{k}(t){=} {\cal T}e^{-i\int_{0}^{t}H_{k}(s)ds}$. For stroboscopic dynamics, i.e., $t=n\tau$ and $n\in \mathbb{Z}^{+}$, $|\Psi(t)\rangle$ can be determined from one the time-period propagator $U(\tau)$. The Floquet theorem allows us to write $U_{k}(n\tau){=}\sum_{\alpha=\pm} e^{-i\mu^{\alpha}_{k}n\tau}|\mu^{\alpha}_{k}\rangle\langle \mu^{\alpha}_{k}|$, where  $|\mu^{\alpha}_{k}\rangle$ are the Floquet modes with $\mu^{\alpha}_{k}$ being the Floquet quasienergies~\cite{quasi}. The $\mu^{\alpha}_{k}$'s are unique only when $\mu^{\alpha}_{k}\in [-\omega/2,\omega/2]$ and repeats such that $\mu^{\alpha,\ell}_{k} = \mu^{\alpha}_{k}+\ell\omega$, $\ell \in \mathbb{Z}$. The above analysis provide us $|\Psi(n\tau)\rangle{=}\sum_k\sum_{\alpha=\pm}e^{-i\mu^{\alpha}_{k}n\tau}|\mu^{\alpha}_{k}\rangle|\langle \mu^{\alpha}_{k}|\Psi_{0}\rangle|^2$. Since $U_{k}\in$ SU(2), it can be written as $U_{k}(\tau)=$
$ 
\begin{pmatrix}
u_{k}(\tau) & -v^{*}_{k}(\tau) \\
v_{k}(\tau) & u_{k}(\tau) 
\end{pmatrix}.
$
Therefore, we can have $\mu^{\pm}_{k} = \pm\frac{\omega}{\pi}\tan^{-1}\sqrt{\frac{1-\Re(u_{k}(\tau))}{1+\Re(u_{k}(\tau))}}$, where $\Re$ stands for the real parts. 


\section{B. Quantum Fisher information in the ground state}
\begin{figure}
\includegraphics[width=0.6\textwidth]{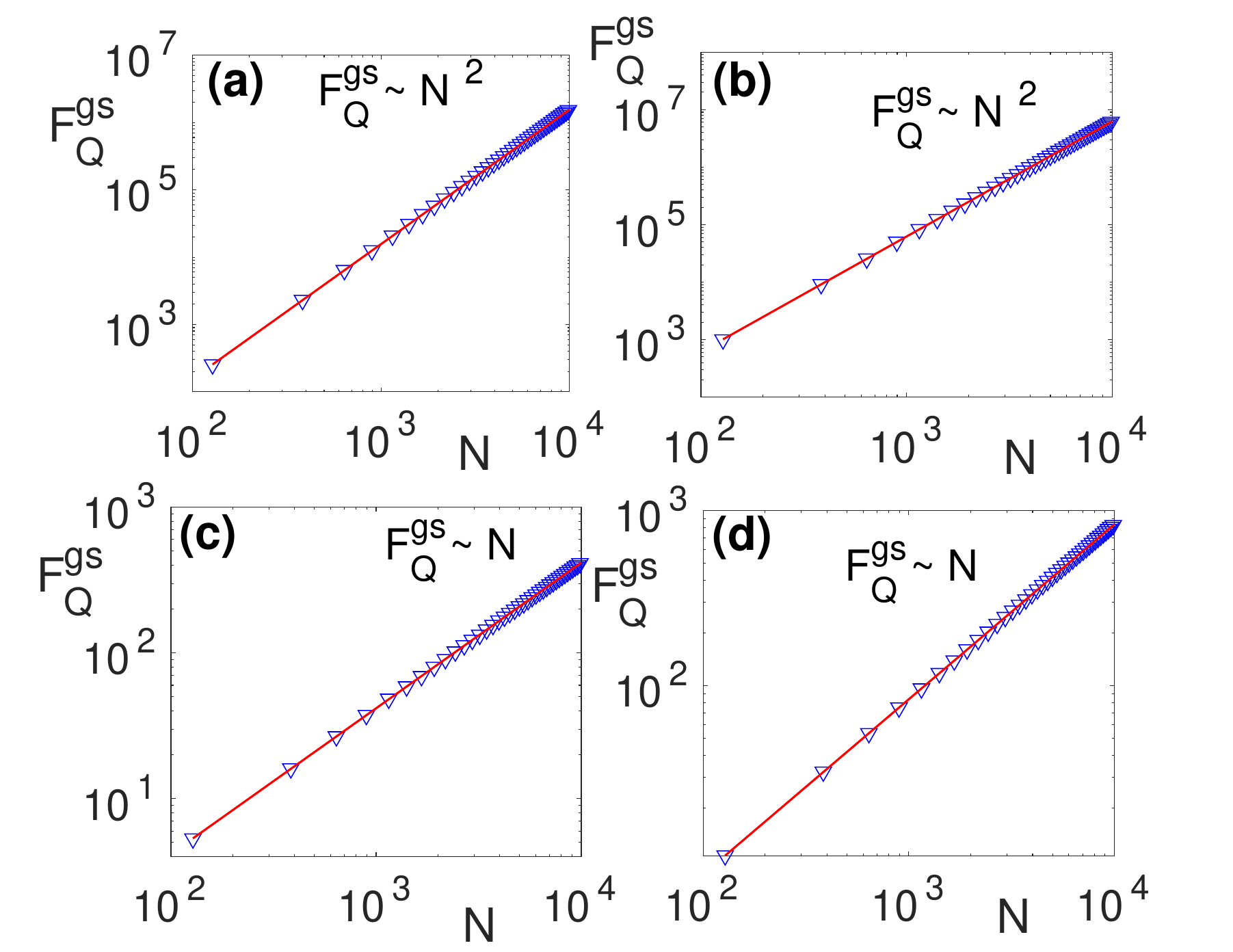}
\caption{Scaling of quantum Fisher information, $F_Q$, as a function of $N$ for (a) $h_0=1,\gamma=0.5$, (b) $h_0=1$, $\gamma=1$, (c) $h_0=0.5$ $\gamma=0.5$, and (d) $h_0=0.5$, $\gamma=1$. }
\label{fig:fig9}
\end{figure}

We analysize the quantum Fisher information, $F^{gs}_Q$, of the full chain in the ground state of the model. We focus on the critical point $h_0$ and obtained $F^{gs}_Q$ as a function of the total system size $N$. Then, we fit the data on the function of the form $\widetilde{F}^{gs}_Q=aN^{\eta}$ such that $\widetilde{F}^{gs}\approx F^{gs}$. The numerical method for obtaining the best fitting function used here is the method of least-sqaure. Once, the best fititng fucntion is obtained, we extract the value of $\eta$ from the fitting function. In Fig.~\ref{fig:fig9}, we peform the fitting and obatained the value of $\eta$ for two different choices of ($h_0,\gamma)$. For $(h_0,\gamma) = (1,0.5)$ and $(h_0,\gamma) = (1,1)$, we find that $\eta=2$ as shown in Figs.~\ref{fig:fig9} (a-b).  This is the celebrated Heisenberg scaling of the quantum Fisher information at the second order phase transitions~\cite{Zanardi2006_1}. In Figs.~\ref{fig:fig9}(c-d), we perform scaling for the value of $h_0$ away from the critical point and the same values of $\gamma$ i.e.,  $(h_0,\gamma) = (0.5,0.5)$ and $(h_0,\gamma) = (0.5,1)$, we obtained $\eta \approx 1$, which known as the standard quantum limit.

\section{C. Steady state correlation functions and quantum Fisher information of a block}
To evaluate the reduced density matrix one needs to compute the correlations functions, ${\cal C}_{i,j}(t){=}\langle \Psi (t)| c^{\dagger}_{i}c_{j}|\Psi(t)\rangle$ and ${\cal I}_{i,j}(t){=} \langle \Psi(t)| c^{\dagger}_{i}c^{\dagger}_{j}|\Psi(t)\rangle$ between the fermionic operators, $i,j=1,\ldots,L$. The density matrix $\rho_L(t)$ is then  characterized by $\Gamma_{ij}(t) {=}\langle \Psi(t)|a_ia_j|\Psi(t)\rangle$, where $a_{2i-1} {=} c_{i}{+}c^{\dagger}_{i}$ and $a_{2i} {=}-i(c_{i}{-}c^{\dagger}_{i})$ are the Majorana operators, such that $\{a_{2i},a_{2j}\}=\{a_{2i-1},a_{2j-1}\}=2\delta_{i,j}$, 
$\{a_{2i},a_{2j-1}\}=0$, and $i,j{=}1,\ldots,L$.   
 The time-dependent reduced density matrix, $\rho_L(t)$, can be obtained from the correlations  ${\cal C}_{ij}$ and ${\cal I}_{ij}$ defined as
\begin{eqnarray}
{\cal C}_{i,j}(t) &= &\frac{2}{N}\sum_{k>0}\cos(k(i-j))\langle d^{\dagger}_{k}d_{k}\rangle_t,\nonumber\\
{\cal I}_{i,j}(t) &= &\frac{2i}{N}\sum_{k>0}\sin(k(i-j))\langle d^{\dagger}_{k}d^{\dagger}_{-k}\rangle_t,
\end{eqnarray}
where, $\langle...\rangle_t$ is the expectation value taken in the time-evolved state. Now we observed that the evolution operator, $U_{k}(t)$, can be expressed in terms of its spectral decomposition as 
\begin{eqnarray}
U_{k}(t) = e^{-i\mu_{+}t}|\mu^{+}_{k}\rangle\langle\mu^{+}_{k}|+e^{-i\mu_{-}t}|\mu^{-}_{k}\rangle\langle\mu^{-}_{k}|.
\end{eqnarray}
With this, we find the  ${\cal C}_{i,j}(t)$ and ${\cal I}_{i,j}(t)$ as

\begin{eqnarray}
{\cal C}_{i,j}(t)&=&\frac{2}{N}\sum_{k>0}\cos(k(i-j))\langle \psi^{0}_{k}| U^{\dagger}_{k}(t)d^{\dagger}_{k}d_{k}U_{k}(t) |\psi^{0}_{k} \rangle\nonumber\\
&=& \frac{2}{N}\sum_{k>0}\cos(k(i-j))\Big[r^{+}_kr^{*+}_k\langle \mu^{+}_k|d^{\dagger}_{k}d_{k}|\mu^{+}_k\rangle\nonumber\\
&+&r^{-}_kr^{*-}_k\langle \mu^{-}_k|d^{\dagger}_{k}d_{k}|\mu^{-}_k\rangle\nonumber\\
&+& e^{i(\mu^{+}_k-\mu^{-}_k)t}r^{+}_kr^{*-}_k\langle\mu^{+}_k|d^{\dagger}_{k}d_{k}|\mu^{-}_k\rangle\nonumber\\
&+& e^{i(\mu^{-}_k-\mu^{+}_k)t}r^{-}_kr^{*+}_k\langle\mu^{-}_k|d^{\dagger}_{k}d_{k}|\mu^{+}_k\rangle
\Big].
\end{eqnarray}

\begin{eqnarray}
{\cal I}_{i,j}(t)&=&\frac{2i}{N}\sum_{k>0}\sin(k(i-j))\langle \psi^{0}| U^{\dagger}_{k}d^{\dagger}_{k}d^{\dagger}_{-k}U_{k} |\psi^{0} \rangle\nonumber\\
&=& \frac{2i}{N}\sum_{k>0}\sin(k(i-j))\Big[r^{+}_kr^{*+}_k\langle \mu^{+}_k|d^{\dagger}_{k}d^{\dagger}_{-k}|\mu^{+}_k\rangle\nonumber\\
&+&r^{-}_kr^{*-}_k\langle \mu^{-}_k|d^{\dagger}_{k}d^{\dagger}_{-k}|\mu^{-}_k\rangle\nonumber\\
&+& e^{i(\mu^{+}_k-\mu^{-}_k)t}r^{+}_kr^{*-}_k\langle \mu^{+}_k|d^{\dagger}_{k}d^{\dagger}_{-k}|\mu^{-}_k\rangle\nonumber\\
&+& e^{i(\mu^{-}_k-\mu^{+}_k)t}r^{-}_kr^{*+}_k\langle\mu^{-}_k|d^{\dagger}_{k}d^{\dagger}_{-k}|\mu^{+}_k\rangle
\Big].
\end{eqnarray}
Note that $|\Psi_{0}(0)\rangle = \otimes_{k}|\psi^{0}_{k}\rangle$ is the initial state and $r_{k}^{\pm}=\langle \psi^{0}_{k}|\mu^{\pm}_{k}\rangle$, describes the overlap of the initial state with that of the Floquet eigenstates.  Taking the limit  $t\to \infty$ and $N\to \infty$, we obtain the correlation functions in the steady-state as
\begin{eqnarray}
{\cal C}^{\infty}_{i,j} &=&\frac{1}{\pi}\int_{0}^{\pi}dk\cos(k(i-j)) \Big[|r^{+}_k|^2\langle \mu^{+}_k|d^{\dagger}_{k}d_{k}|\mu^{+}_k\rangle\nonumber\\
&+&|r^{-}_k|^{2}\langle \mu^{-}_k|d^{\dagger}_{k}d_{k}|\mu^{-}_k\rangle\Big],
\end{eqnarray}
\begin{eqnarray}
{\cal I}^{\infty}_{i,j} &=&\frac{i}{\pi}\int_{0}^{\pi}dk\sin(k(i-j)) \Big[|r^{+}_k|^2\langle \mu^{+}_k|d^{\dagger}_{k}d^{\dagger}_{-k}|\mu^{+}_k\rangle\nonumber\\
&+&|r^{-}_k|^{2}\langle \mu^{-}_k|d^{\dagger}_{k}d^{\dagger}_{-k}|\mu^{-}_k\rangle\Big],
\end{eqnarray}
where we replace the summation by integration. The non-zero elements of the $\Gamma$ matrix, therefore,  are given by 
\begin{eqnarray}
\Gamma_{2i-1,2j-1} &=& \delta_{i,j}+2i\Im\Big({\cal C}_{i,j}+{\cal I}_{i,j}\Big)\nonumber\\
\Gamma_{2i-1,2j} &=& i\delta_{i,j}-2i\Re\Big({\cal C}_{i,j}-{\cal I}_{i,j}\Big)\nonumber\\
\Gamma_{2i,2j-1} &=& -i\delta_{i,j}+2i\Re\Big({\cal C}_{i,j}+{\cal I}_{i,j}\Big)\nonumber\\
\Gamma_{2i,2j} &=& \delta_{i,j}+2i\Im\Big({\cal C}_{i,j}-{\cal I}_{i,j}\Big).
\end{eqnarray}
Once we obtain the matrix $\Gamma$, we can get the quantum Fisher information $F_Q$ of a block of size $L$~\cite{Carollo2019_1,UM2020_1} as

\begin{eqnarray}
F_Q(t) = \sum\limits_{r,s=1}^{2L}\frac{\langle r|\partial_{h_0}\Gamma|s\rangle\langle s|\partial_{h_0}\Gamma|r\rangle}{(1-\lambda_r \lambda_s)}.
\label{eq:fihser_gamma}
\vspace{-0.5cm}
\end{eqnarray}
Here, $\Gamma{=}\sum_{r=1}^{2L} \lambda_r|r\rangle\langle r|$ is the spectral decomposition of $\Gamma$ and $\partial_{h_0}\Gamma{=}\frac{\partial \Gamma}{\partial h_0}$. The above formula can show singular behavior at $\gamma_{r}=\gamma_{s}=\pm 1$. It is shown that the abobe singularity can be removable~\cite{singularFisher}.

\section{D. Role of anisotropy in the QFI}
\begin{figure}
\includegraphics[width=0.6\textwidth]{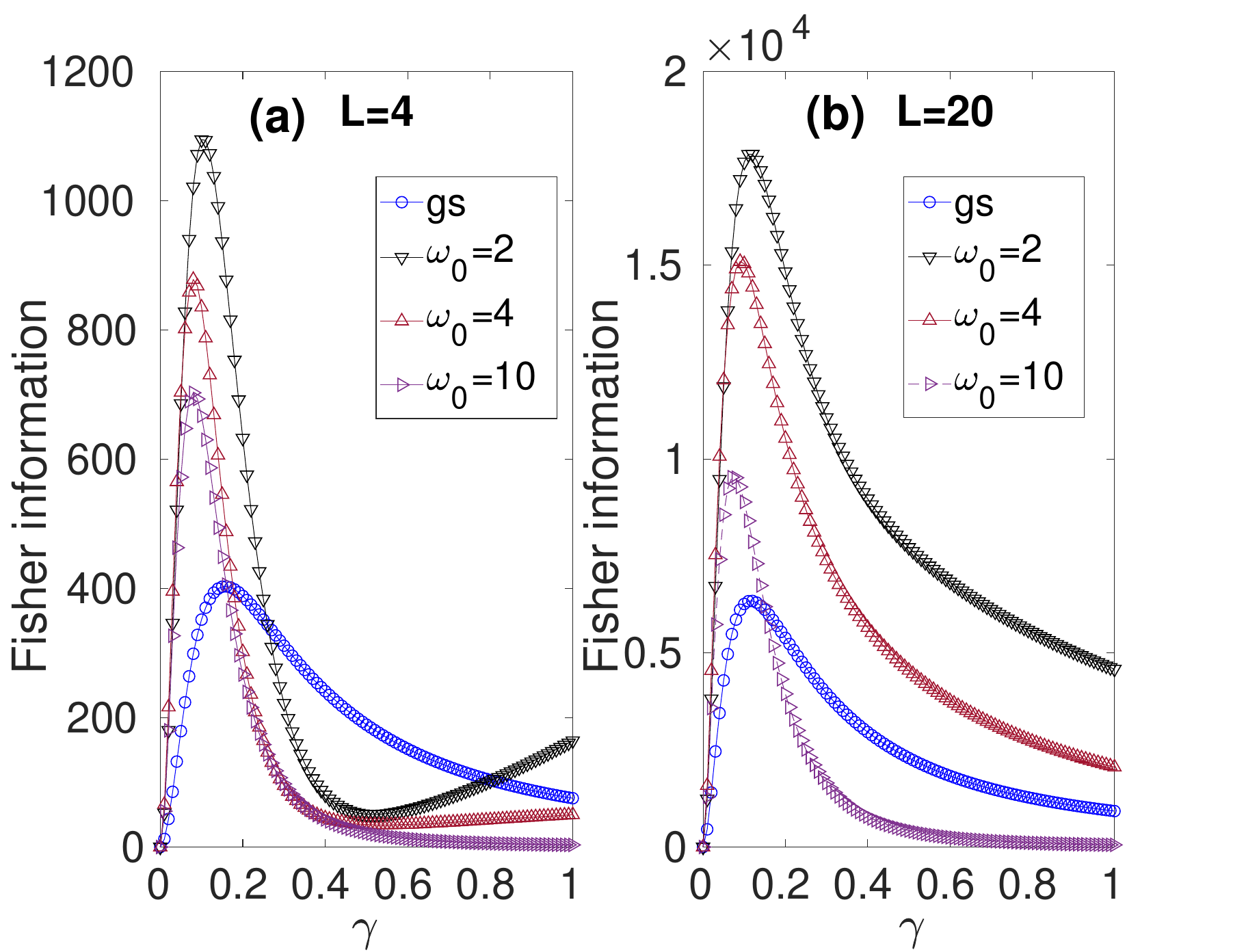}
\caption{$F^{gs}_Q$ and $F^{ss}_{Q}$ as a function of $\gamma$ for (a) $L=4$ and (b) $L=20$. Here  $h_0=1$, $h_1=1.5$, and  $N=6000$. }
\label{fig:fig8}
\end{figure}

In Fig.~\ref{fig:fig8}, we plot $F^{ss}_{Q}$ as a function of anisotropy parameter $\gamma$ for $h_0/J=1$ for different frequencies, $\omega$.  In Fig.~\ref{fig:fig8}(a), the $F^{ss}_{Q}$ is for $L=4$ and in Fig.~\ref{fig:fig8}(b), it is for $L=20$. It can be seen that as  we increase $|\gamma|$, $F^{ss}_Q$ increases monotonically with $\gamma$. The $F^{ss}_{Q}$ shows a peak at some value of $\gamma$ which depends on the $\omega$. It then decreases with the $|\gamma|$. The steady-state $F^{ss}_{Q}$ is greater than the ground state $F^{gs}_Q$ for a range of $\gamma$. In fact for certain frequencies and block size $L$, the $F^{ss}_{Q}>F_Q$ for all $\gamma$.

\begin{figure}
\includegraphics[height=0.4\textwidth]{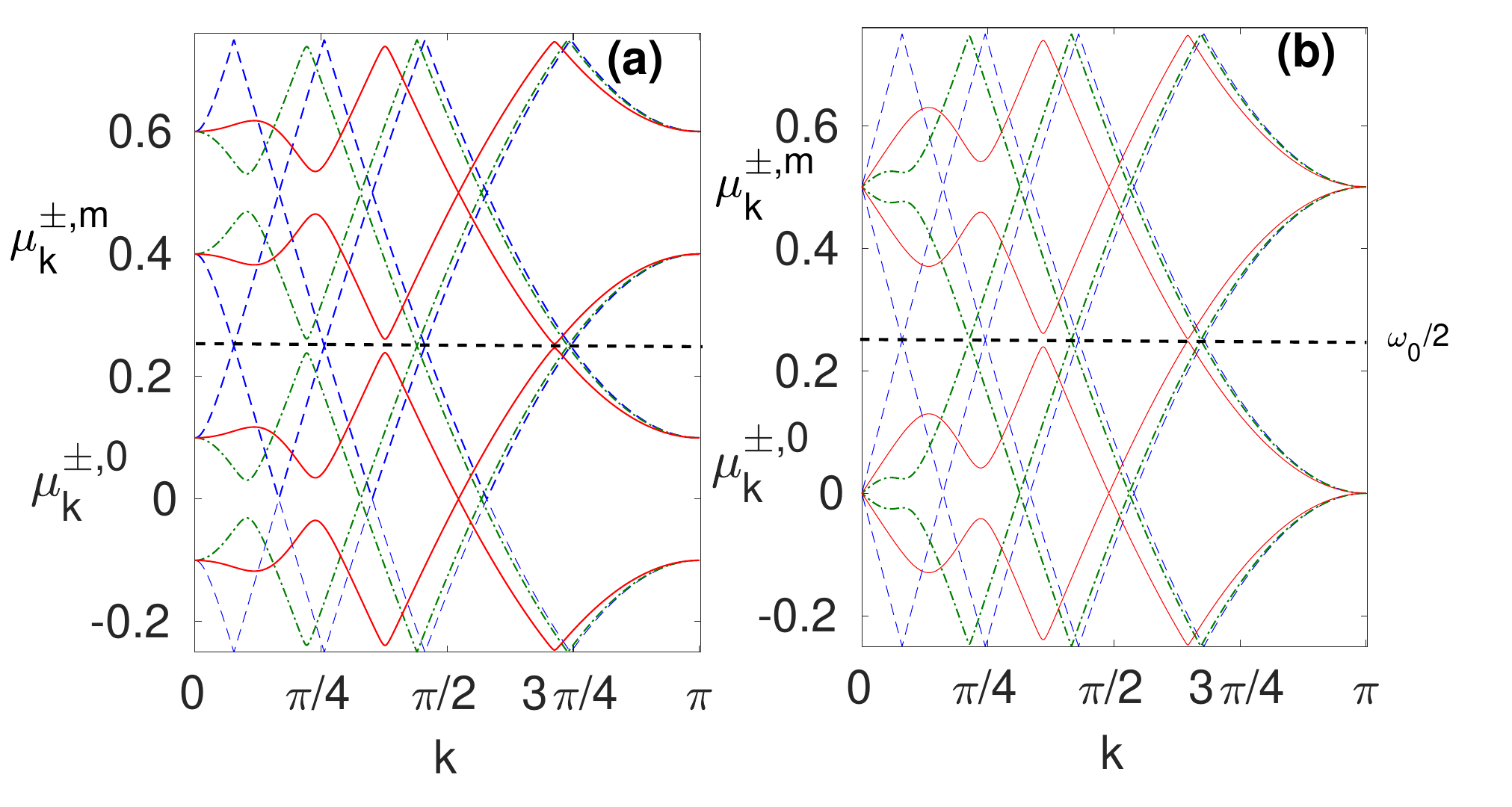}
\caption{Quasienergies $\mu_{k}^{\pm}$ as a function of $k$ for (a) $h_0=0.9$ and (b) $h_0=1$. The different curves in the panel are for different $h_1$: $h_1=0.0$ (blue dashed) $h_1=0.5$ (green dot dashed), and $h_1=1$ (solid red lines). Here $\omega=0.5$, $h_0=1$, $\gamma=1$, and $m=1$.  }
\label{fig:fig7}
\end{figure}

\section{E. Floquet theory and  Floquet resonances}
In this sections,  we give analytical arguments for the occurrence of Floquet resonances for all values of $h_1$. For a  time-periodic many-body Hamiltonian, $H(t+\tau)=H(t)$, with periodicity $\tau=2\pi/\omega$, the solution of the Schr\"{o}dinger  equation follows from the Floquet theorem. The Floquet theorem gives an ansatz of the form $|\Psi(t)\rangle=\sum_{\alpha} c^{\alpha}e^{-i\mu^{\alpha}t}|\Phi^{\alpha}(t)\rangle$. Here $c^{\alpha}\in \mathbb{C}$, $\mu^{\alpha}$ are quasienergies, and $|\Phi^{\alpha}(t)\rangle$ are Floquet modes. 
Substituting the ansatz to the Schr\"{o}dinger equation, $H(t)|\Psi(t)\rangle = -i\hbar \frac{\partial |\Psi(t)\rangle}{\partial t}$, the Floquet modes satisfy
\begin{equation}
\Big(H(t)-i \frac{\partial }{\partial t}\Big)|\Phi^{\alpha}(t)\rangle=\mu^{\alpha}|\Phi^{\alpha}(t)\rangle.
\end{equation} 
It can be noted that $|\Phi^{\alpha}(t+\tau)\rangle$ is also a solution of the above equation with quasienergy $\mu^{\alpha}$, therefore, we have $|\Phi^{\alpha}(t+\tau)\rangle = |\Phi^{\alpha}(t)\rangle$. 
For the model, we consider propagator $U(t) = {\cal T }e^{-i\int^{t}_{0}H(s) ds}$ so that $U(t+\tau)|\Psi(t)\rangle=|\Psi(t+\tau)\rangle$. Using the ansatz for $|\Psi(t)\rangle$ in the above equation, we have $U(t+\tau)e^{-i\mu^{\alpha}t}|\Phi^{\alpha}(t)\rangle = e^{-i\mu^{\alpha}(t+\tau)}|\Phi^{\alpha}(t+\tau)\rangle =e^{-i\mu^{\alpha}(t+\tau)}|\Phi^{\alpha}(t)$. This shows that the Floquet modes are eigenvectors of the one time-period propagator $U(\tau)$ and quasienergies $\mu^{\alpha}$ are its eigenvalues. We numerically diagonalize the one time-period propagator $U_{k}(\tau)$ for each $2\times 2$ $k$-space and obtained the Floquet modes $|\mu^{\pm}_{k}\rangle$ and quasienergies $\mu^{\pm}_{k}$.

As it has been shown in the main text, Figs.~3(a-b), that there is a peak in $F^{ss}_{Q}$ for certain $h_0$ for a fixed  $\omega$. In Figs.~3(c-d), we find that the peaks occurs at the point where Floquet gap vanishes. The peaks  have also been observed before in the other quantities~\cite{Pd_steady_state_1,ent2_1,ent3_1}. Here, we explain the occurrence of Floquet resonance in the system observed in the QFI following Ref.~\cite{Pd_steady_state_1}. The Floquet modes $|\mu^{\pm}_{k}\rangle$ are defined upto a periodic phase, i.e,  $|\mu^{\pm, \ell}_{k}(t)\rangle = e^{i\omega \ell t}|\mu^{\pm}_{k}\rangle$, where $\ell$ is an integer. The new Floquet quasienergies are, therefore, shifted as $\mu_{k}^{\pm, \ell} = \mu^{\pm}_{k}\pm \ell \omega$. Thus, the quasienergies are uniquely defined upto a translation of an integer multiple of $\omega$. It is feasible, therefore, to defined the quasienergies within the first Brillouin zone $[-\omega/2,  \omega/2]$. In the absence of driving, $h_1\to 0$, $U_k(\tau,0) = e^{-iH_k\tau}$. Thus, for the critical sensing, i.e., for $h_0=1$, the Floquet quasienergies are the eigenvalues of the critical $H_k$. Thus, we have $\mu_{k}^{\pm,\ell}=\pm 2\gamma\sin(k/2)+l \omega$. The degeneracy condition translate into $4\gamma\sin(k/2)=p\omega$, where $p\in \mathbb{N}$. This is known as the $n$-photon resonance condition. For finite $h_1$, and for $h_0=0.9$ (different from the value where $F^{ss}_Q$ is peaked), the quasienergies are always gapped for all $k$, as can be seen from Fig.~\ref{fig:fig7}(a). Thus, there will not be any Floquet resonance for $h_0=0.9$. On the other hand, for $h_1\neq 0$ and $h_0=1$, the quasienergies open a gap at all $k$ except at $k=0,\pi$, as can be seen from Fig.~\ref{fig:fig7}(b). In Fig.~\ref{fig:fig7}(b), we plot the quasienergies for various $h_1$ in the range $[-3\omega/2,\omega/2]$ for $\omega=0.5$. It can be seen from the plot that by translating $\mu_{k}^{\pm}\to \mu_{k}^{\pm}+\omega$, we get the same quasienergies within the range $[-\omega/2,\omega/2]$. Moreover, it can be seen from the plot that as $h_1$ increases from $h_1=0$ to $h_1=1$, there is a gap in the quasienergies except at $k=0,\pi$.  To obtained the degeneracy condition for $h_1\neq 0$, one can first change the system in a rotating frame via a time-dependent transformation ${\cal V}(t)=\exp\Big(-ih_1\frac{\cos(\omega t)}{\omega}\sigma^z\Big)$. This gives a transformed Hamiltonian ${\cal H}_{k}(t) = {\cal V}_{k}^{\dagger}(t)H_{k}(t){\cal V}_{k}(t)-i{\cal V}^{\dagger}(t)\dot{{\cal V}}(t)$. The new Hamiltonian is ${\cal H}_{k} = (h_0-J\cos(k))\sigma^{z}+\gamma \sigma^{y} \sin(k)\exp\Big(2ih_1\frac{\cos(\omega t)}{\omega}\Big)$. Now for the modes for which the amplitude $h_1$ has small effect on quasienergies can be neglected. Therefore, the second term of ${\cal H}_k$ will not contribute in determining the Floquet resonances. The Floquet Hamiltonian ($H^{F}_{k}$), then, can be constructed as
\begin{eqnarray}
e^{-iH^{F}_k\tau}={\cal T} e^{-i\int_{0}^{\tau} H_{k}(t)dt}
\end{eqnarray} 
The quasienergies, i.e., the eigenvalues of $H^{F}_{k}$, are given by $\mu_{k=0}^{\pm,\ell} = \pm (h_0- h_c) + \ell \omega$ and $\mu_{k=\pi}^{\pm,\ell} = \pm (h_0+ h_c) + \ell \omega$ upto the translation of an integer multiple of $\omega$. It can be noted that $\pm (h_0- h_c) + \ell \omega$ and $\pm (h_0+ h_c) + \ell \omega$  are also the eigenvalues of  ${\cal H}_{k=0}(t)$ and ${\cal H}_{k=\pi}(t)$, respectively. The degeneracy condition of quasienergies gives $2(h_0\pm h_c)=q\omega$ with $q\in \mathbb{N}$, which is equivalent to $E_{k=0,\pi}(t=0)=q\omega$.

\begin{figure}
\includegraphics[width=0.6\textwidth]{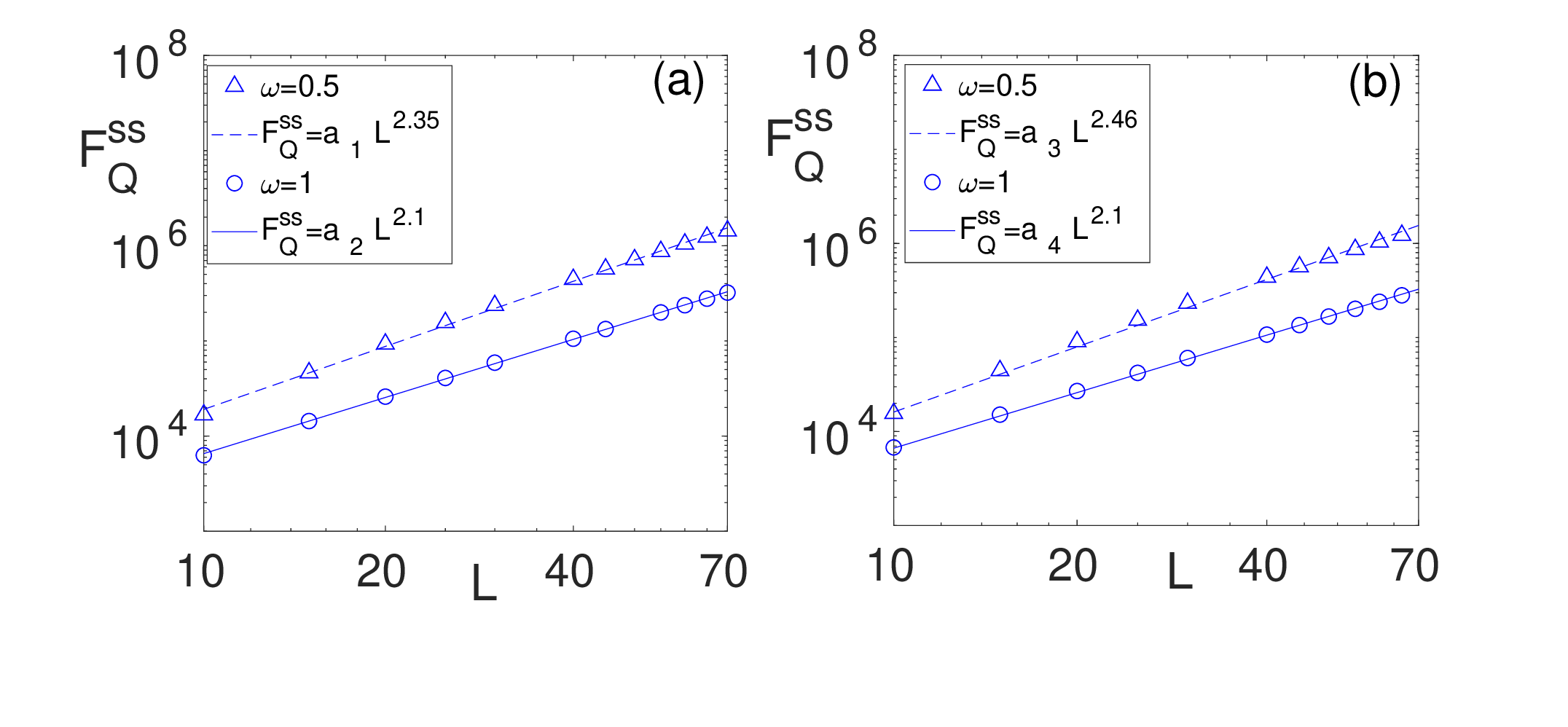}
\caption{Fitting of the $F^{ss}_{Q}$ as a function of $L$ with fitting function $F^{ss}_{Q}=aL^{\eta}$ for different initial states: (a) $|\psi_1\rangle = |0\otimes 0\otimes\ldots\otimes 0\rangle$=$|0\rangle^{\otimes N}$ and (b) $|\psi_2\rangle=|+ \otimes +\otimes\ldots\otimes+\rangle$ =$|+\rangle^{\otimes N}$, where $|+\rangle=\frac{1}{\sqrt{2}}(|0\rangle+|1\rangle)$. The fitting coefficients are: $a_1 =4.208$, $a_2 =3.74,$ $a_3 =3.76$, and $a_4=3.75$. Here $h_0=1$,  $\gamma=1$, and $h_1=1.5$. }
\label{fig:S4}
\end{figure}

\section{F. Classical Fisher Information}
The classical Fisher information $F_C$ with respect to the parameter $h_0$ is defined as 
\begin{equation}
F_C=\sum_{i}\frac{1}{p_i}\Big(\frac{\partial p_i}{\partial h_0}\Big)^2,
\label{eq:FC}
\end{equation}
where $p_i$'s are the probabilities of a positive operator valued measurement (POVM) such that each $p_i>0$ and $\sum_{i}p_i=1$.  For the present case, we consider a simple, though sub-optimal, measurement which is independent of $h_0$. The measurement is the block magnetization along $x$-direction for a block of size $L$. For a block of size $L$, the global magnetization takes $L+1$ outcomes from  $O_1=+L$ (when all the qubits are $|+\rangle$), $O_2=L-2$ (when except one qubit the rest are in the state $|+\rangle$) until $O_{L+1}=-L$ (when all the qubits are $|-\rangle$). Here $|\pm\rangle = \frac{1}{\sqrt{2}}(0\rangle\pm|1\rangle$.
Each of the outcomes $O_r$ has probability $p_r$ of occurrence. Then one can use Eq.~(\ref{eq:FC}) to get the corresponding classical Fisher information $F_C$.

\section{G. Role of the initial state}
In this section, we present results for the scaling of the $F^{ss}_{Q}$ for the two initial states (i) $|\psi_1\rangle = |0\otimes 0\otimes\ldots\otimes 0\rangle$=$|0\rangle^{\otimes N}$ and (ii) $|\psi_2\rangle=|+ \otimes +\otimes\ldots\otimes+\rangle$ =$|+\rangle^{\otimes N}$, where $|+\rangle=\frac{1}{\sqrt{2}}(|0\rangle+|1\rangle)$ and $\sigma^{z}|0\rangle=|0\rangle$ and $\sigma^{z}|1\rangle=-|1\rangle$. In Fig.~\ref{fig:S4}, we show scaling of the $F^{ss}_{Q}$ with subsystem size $L$ at $h_0=1$ for $|\psi_1\rangle$ in Fig.~\ref{fig:S4}  (a) and $|\psi_2\rangle$ in Fig.~\ref{fig:S4} (b). From the obtained scaling exponent $\eta$ using method of least square fitting, we found that the scaling exponent is almost same for all the initial states considered. Thus, we can conclude that the scaling behavior is independent of the initial state and depend on the occurrence of vanishing Floquet gap as discussed in the main text.

\end{document}